\documentclass[%
 aip,
 amsmath,amssymb, 
 reprint,%
floatfix,
apl
]{revtex4-1}

\usepackage{graphicx}
\usepackage{dcolumn}
\usepackage{bm}
\usepackage{soul}
\usepackage{xcolor}

\usepackage[utf8]{inputenc}
\usepackage[T1]{fontenc}
\usepackage{mathptmx}
\usepackage{etoolbox}
\usepackage{float}
\usepackage[colorlinks=true,allcolors=teal]{hyperref}

\usepackage[sectionbib]{bibunits}

\makeatletter
\def\@email#1#2{%
 \endgroup
 \patchcmd{\titleblock@produce}
  {\frontmatter@RRAPformat}
  {\frontmatter@RRAPformat{\produce@RRAP{*#1\href{mailto:#2}{#2}}}\frontmatter@RRAPformat}
  {}{}
}%
\makeatother

\defaultbibliographystyle{apsrev4-1} 
\defaultbibliography{bib}

\begin{document}
\title[Loss tangent fluctuations due to two-level systems in superconducting microwave resonators]{Loss tangent fluctuations due to two-level systems in superconducting microwave resonators}
\author{André Vallières}
\affiliation{Graduate Program in Applied Physics, Northwestern University, Evanston, IL 60208, USA\looseness=-1}
\affiliation{Fermi National Accelerator Laboratory, Batavia, IL 60510, USA}
\email{andrevallieres2027@u.northwestern.edu}

\author{Megan E. Russell}
\affiliation{Department of Physics and Astronomy, Northwestern University, Evanston, IL 60208, USA\looseness=-1}

\author{Xinyuan You}
\affiliation{Fermi National Accelerator Laboratory, Batavia, IL 60510, USA}

\author{David A. Garcia-Wetten}
\affiliation{Department of Materials Science and Engineering, Northwestern University, Evanston, IL 60208, USA\looseness=-1}

\author{Dominic P. Goronzy}
\affiliation{Department of Materials Science and Engineering, Northwestern University, Evanston, IL 60208, USA\looseness=-1}

\author{Mitchell J. Walker}
\affiliation{Department of Materials Science and Engineering, Northwestern University, Evanston, IL 60208, USA\looseness=-1}

\author{Michael J. Bedzyk}
\affiliation{Department of Physics and Astronomy, Northwestern University, Evanston, IL 60208, USA\looseness=-1}
\affiliation{Department of Materials Science and Engineering, Northwestern University, Evanston, IL 60208, USA\looseness=-1}

\author{Mark C. Hersam}
\affiliation{Department of Materials Science and Engineering, Northwestern University, Evanston, IL 60208, USA\looseness=-1}
\affiliation{Department of Chemistry, Northwestern University, Evanston, IL 60208, USA\looseness=-1}
\affiliation{Department of Electrical and Computer Engineering, Northwestern University, Evanston, IL 60208, USA\looseness=-1}

\author{Alexander Romanenko}
\affiliation{Fermi National Accelerator Laboratory, Batavia, IL 60510, USA}

\author{Yao Lu}
\affiliation{Fermi National Accelerator Laboratory, Batavia, IL 60510, USA}

\author{Anna Grassellino}
\affiliation{Fermi National Accelerator Laboratory, Batavia, IL 60510, USA}

\author{Jens Koch}
\affiliation{Department of Physics and Astronomy, Northwestern University, Evanston, IL 60208, USA\looseness=-1}
\affiliation{Center for Applied Physics and Superconducting Technologies, Northwestern University, Evanston, IL 60208, USA\looseness=-1}

\author{Corey Rae H. McRae}
\affiliation{Electrical, Computing and Energy Engineering Department, University of Colorado, Boulder, CO 80309, USA\looseness=-1}
\affiliation{Department of Physics, University of Colorado, Boulder, CO 80309, USA}
\affiliation{National Institute of Standards and Technology, Boulder, CO 80305, USA}

\date{18 March 2025}

\begin{abstract}
Superconducting microwave resonators are critical to quantum computing and sensing technologies. Additionally, they are common proxies for superconducting qubits when determining the effects of performance-limiting loss mechanisms such as from two-level systems (TLS). The extraction of these loss mechanisms is often performed by measuring the internal quality factor $Q_i$ as a function of power or temperature. In this work, we investigate large temporal fluctuations of $Q_i$ at low powers over periods of 12 to 16 hours (relative standard deviation $\sigma_{Q_i}/Q_i = 13\%$). These fluctuations are ubiquitous across multiple resonators, chips and cooldowns. We are able to attribute these fluctuations to variations in the TLS loss tangent due to two main indicators. First, measured fluctuations decrease as power and temperature increase. Second, for interleaved measurements, we observe correlations between low- and medium-power $Q_i$ fluctuations and an absence of correlations with high-power fluctuations. Agreement with the TLS loss tangent mean is obtained by performing measurements over a time span of a few hours. We hypothesize that, in addition to decoherence due to coupling to individual near-resonant TLS, superconducting qubits are affected by these observed TLS loss tangent fluctuations.
\end{abstract}

\maketitle

\begin{bibunit}
In superconducting circuits, microwave resonators are implemented for many purposes, including qubit readout,~\cite{Blais_2004,Wallraff_2004} quantum memory,\cite{Hofheinz_2009} and quantum sensing.~\cite{Mazin_2004} Superconducting microwave resonators have also become essential in characterizing losses in superconducting circuits with the goal of making higher coherence qubits \cite{Martinis_2005, Woods_2019, Pappas_2011, Dunsworth_2018, McRae_dielectric_2020, Muller_2019}. Measuring the intrinsic quality factors $Q_i$ of resonators at different powers and temperatures enables the separation of the total loss into contributions from two-level systems (TLS), equilibrium quasi-particles (QP), and other loss sources such as radiation or coupling to parasitic modes \cite{McRae_2020, Crowley_2023}. 

Although there has been significant effort to characterize frequency fluctuations in resonators \cite{Gao_2007, Kumar_2008, Barends_2009, Lindström_2011, Gao_2011, DeGraaf_2018, Burnett_2018, Niepce_2021}, the same cannot be said for fluctuations in dissipation, especially for distributed-element resonators over long time scales. While Ref.~\citenum{Béjanin_2022} measures resonator dissipation over long time scales, we expand this line of inquiry to describe how $Q_i$ fluctuations arise from variations in the TLS loss tangent. This is evidenced by our observations of fluctuations decreasing at higher powers and temperatures, consistent with TLS saturation. We report quality factor fluctuations at low power with relative standard deviation $\sigma_{Qi}/Q_i$ of 13\% in distributed-element superconducting resonators across multiple resonators, chips and cooldowns. We analyze the impacts of these fluctuations on the results of common methods of TLS loss tangent extraction.

In the case of qubits, it is becoming standard to report statistics of $T_1$ rather than a single value since it is known that the relaxation time $T_1$ can substantially fluctuate over time \cite{Klimov_2018,  Burnett_2019, Béjanin_2021, Carroll_2022, Zhu_2024}. Similarly, we demonstrate with resonators that rapidly measuring over a few hours appropriately captures the TLS loss tangent mean and standard deviation, while performing time-averaged measurements for the same time frame represents well the mean but does not capture the width of the distribution.

\begin{figure*}[t!]
    \includegraphics[width=\hsize]{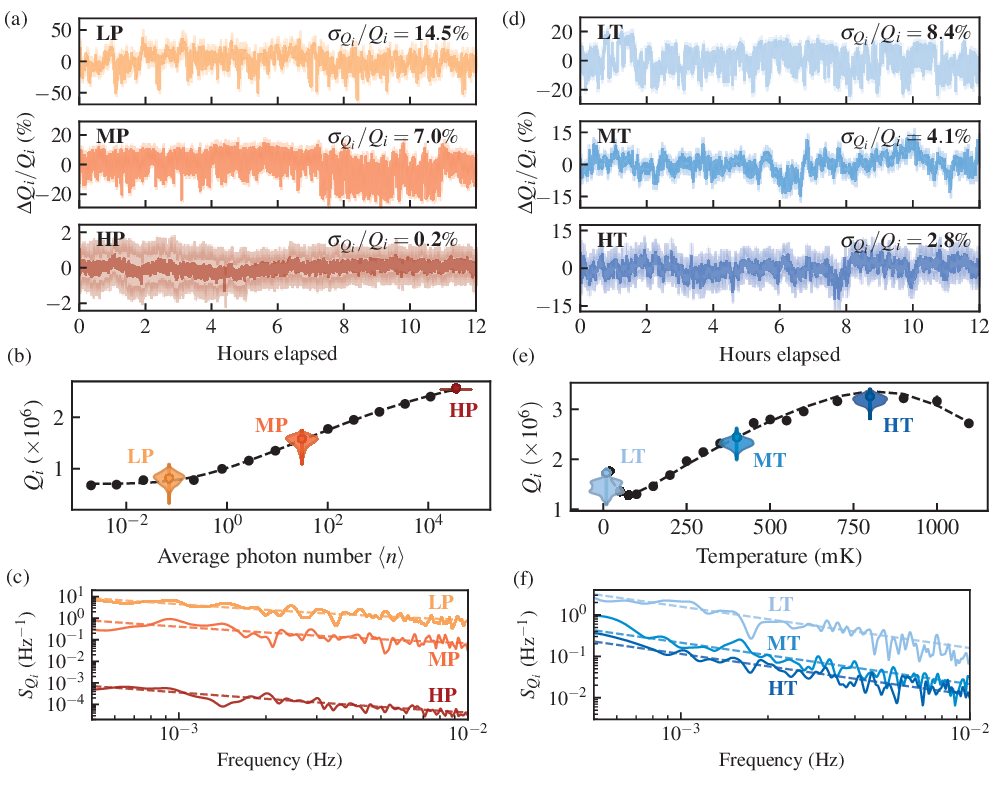}
    \caption{Internal quality factor fluctuations with varying power (a-c) and temperature (d-f). Measurements for (a-c) were performed at base temperature ($\sim$10\,mK) and at three different powers: low (LP, -75\,dBm), medium (MP, -60\,dBm), and high (HP, -25\,dBm). (a) Time traces of the relative deviation from the mean $\Delta Q_i/Q_i$, showing decreasing fluctuations with increasing power, consistent with TLS saturation. The pale regions around the data represent fitting uncertainty. (b) The three powers are represented on the TLS loss model curve with a fit to Eq.~\eqref{eq:TLS_loss} shown as a dashed line. The violins show the density curves of the time traces in (a) and additionally provide a visual comparison of the fluctuation magnitudes at different powers. (c) Low-frequency noise spectra $S_{Q_i}$ further demonstrating a decrease in fluctuations with increasing power. Notably, we observe a difference of four orders of magnitude between low and high powers. Fits to $1/f$ spectra are also shown as dashed lines to serve as guides to the eye. (d-f) Quality factor fluctuations at a power of -55\,dBm at the VNA and three different temperatures: low (LT, 10\,mK), medium (MT, 400\,mK), and high (HT, 800\,mK). In (d), we show the time traces of the relative deviation from the mean at these temperatures. We observe decreasing fluctuations at higher temperatures, also consistent with TLS saturation. The pale regions around the data represent fitting uncertainty. We choose the temperatures in relation to TLS saturation levels as described by the model in Ref.~\citenum{Crowley_2023}; the fit is shown as a dashed line in (e). The TLS-dominated and QP-dominated regimes are shown below and above $\sim$800\,mK, respectively. The violins again correspond to the density curves for the traces in (d).  The vertical lines at each black point correspond to fitting uncertainty of individual measurements. (f) Low-frequency noise spectra $S_{Q_i}$ for the three temperatures, further demonstrating decreasing fluctuations with increasing temperature. Fits to $1/f$ spectra are again represented as dashed lines to serve as guides to the eye. We note that in this figure, we used two representative resonators: one for the power dependence results and another for the temperature dependence results.}
    \label{fig:fluctuations-vs-power-temp}
\end{figure*}

All measured resonators are Nb on Si coplanar waveguide quarter-wavelength resonators inductively coupled to a feedline (see Appendix A for more information). The measurement chain consists of a vector network analyzer (VNA) to perform transmission ($S_{21}$) measurements across the samples, a Josephson parametric amplifier (JPA) used at lower powers, and suitable attenuators, filters, isolators and additional amplifiers for achieving desirable signal quality in the 4\,GHz to 8\,GHz range (see Appendix B). The transmitted signal is analyzed using a circle fit routine \cite{Khalil_2012, Baity_2024} to extract $Q_i$ and the resonance frequency $f_r$ following
\begin{equation}
    S_{21}\left(f\right) = A e^{-i 2 \pi f \tau} \left(1 - \frac{(Q/|Q_c|) e^{i \phi}}{1 + 2 i Q (f/f_r - 1)} \right),
    \label{eq:S21-hanger}
\end{equation}
where $A$ is the (complex) attenuation of the line and $\tau$ is the line delay. The internal quality factor $Q_i$ is obtained from the loaded $Q$ and complex coupling quality factor $Q_c = |Q_c| e^{-i \phi}$ using $1/Q_i = 1/Q - \cos\phi/|Q_c|$, where the angle $\phi$ accounts for the impedance mismatch between the feedline and the resonator. Obtaining $Q_i$ this way, we are able to independently measure the fluctuations of $f_r$ and $Q_i$. Reported fitting uncertainties in the text correspond to the 68\% confidence interval.

In Fig.~\ref{fig:fluctuations-vs-power-temp}, we explore the dependence of $Q_i$ fluctuations on power and temperature. We observe a systematic decrease of fluctuations at higher powers and temperatures. This aligns with the idea that the fluctuations are caused by TLS: namely when TLS become saturated at high powers and temperatures, they contribute less to the total loss~\cite{Neill_2013, Béjanin_2022}. We note that these results hold for all resonators measured as part of this study (see Appendix H).

The analysis of the power dependence is based on a TLS saturation model~\cite{Gao_Thesis_2008,Burnett_2018,Muller_2019,McRae_2020}, according to which:
\begin{equation}
\begin{split}
    \frac{1}{Q_i} &= \frac{1}{Q_{\mathrm{TLS}}(n,T)} + \frac{1}{Q_{\text{PI}}} \\
    &= F \delta^0_{\mathrm{TLS}} \frac{\tanh{\frac{\hbar \omega_r}{2 k_B T}}}{\left( 1 + \frac{\langle n \rangle}{n_c}\right)^\beta} + \frac{1}{Q_{\text{PI}}}.
    \label{eq:TLS_loss}
\end{split}
\end{equation}
Here, $F$ is the filling factor, $\delta^0_{\mathrm{TLS}}$ ($Q_{\mathrm{TLS}}(n,T)$) the intrinsic TLS loss tangent (quality factor), $\omega_r$ the angular resonance frequency of the resonator, $k_B$  the Boltzmann constant, $T$ the temperature, $\langle n \rangle$ the average photon number in the resonator, $n_c$ the critical photon number, $\beta$ an empirical parameter describing TLS interaction, and $1/Q_{\text{PI}}$ represents the power-independent loss. This model is used to obtain the effective TLS loss tangent $F \delta^0_{\mathrm{TLS}}$ from a fit of $Q_i$ versus $\langle n \rangle$. It also informs our choice of powers to investigate fluctuations at different levels of TLS saturation. Details on the relationship between the power from the VNA and $\langle n \rangle$ is given in the appendix.

Fig.~\hyperref[fig:fluctuations-vs-power-temp]{\ref*{fig:fluctuations-vs-power-temp}(a)} shows 12-hour time traces of $Q_i$ for measurements at low (LP), medium (MP), and high (HP) powers, corresponding to -75\,dBm, -50\,dBm and -20\,dBm at the VNA, respectively. We observe considerable fluctuations, especially at low power where $Q_i$ varies by as much as 37\% from the average during the 12 hours. We further illustrate the fluctuations' magnitude in Fig.~\hyperref[fig:fluctuations-vs-power-temp]{\ref*{fig:fluctuations-vs-power-temp}(b)} relative to the TLS loss model curve (S curve).  Additionally, we characterize the fluctuations in the frequency domain by plotting the spectral density of the normalized $Q_i$ time series $(Q_i(t) - \langle Q_i\rangle)/\langle Q_i \rangle$ estimated using Welch's method~\cite{Welch_note}. Here, $\langle Q_i \rangle$ refers to the average value of the time trace. We observe that fluctuations at high power are reduced by as much as four orders of magnitude compared to low power.

The temperature dependence measurements were performed and analyzed in a similar fashion. We choose temperatures relative to TLS saturation levels based on our measurements of $Q_i$ and the model described in Ref.~\citenum{Crowley_2023}. This model explains $Q_i(T)$ by accounting for temperature-dependent TLS saturation and QP loss. Fig.~\hyperref[fig:fluctuations-vs-power-temp]{\ref*{fig:fluctuations-vs-power-temp}(d)} shows 12-hour time traces of $Q_i$ at low (LT), medium (MT), and high (HT) temperatures, corresponding to 10\,mK, 400\,mK, and 800\,mK, respectively. We observe decreasing fluctuations with increasing temperature, with an order of magnitude difference between LT and HT as shown in Fig.~\hyperref[fig:fluctuations-vs-power-temp]{\ref*{fig:fluctuations-vs-power-temp}(f)}. These results, along with those from power dependence measurement, support the claim that the large fluctuations at low power and temperature are caused by TLS.

\begin{figure}[t!]
\includegraphics[width=\hsize]{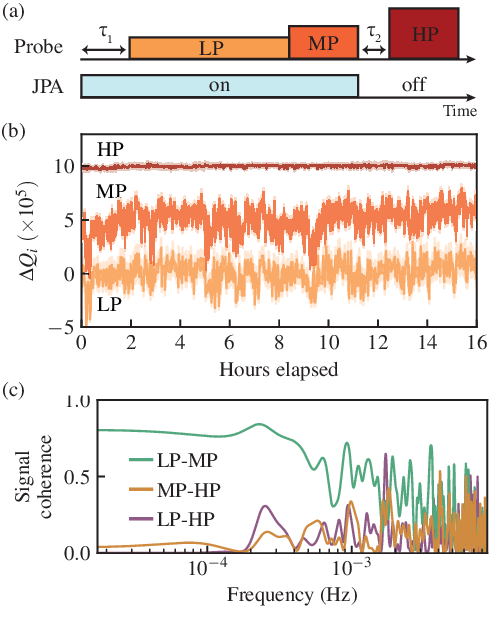}
    \caption{Interleaved power measurements of the $Q_i$ deviation from the average $\Delta Q_i = Q_i - \langle Q_i \rangle$ at base temperature ($\sim$10\,mK). (a) Measurement sequence for interleaved measurements, with the JPA turned off during high-power measurements. Idling times of $\tau_1 = 3$\,s and $\tau_2 = 0.5$\,s were found to be sufficient to ensure that the JPA is stable when performing measurements. This sequence is repeated over 16 hours. (b) Time traces at low (LP, -75\,dBm), medium (MP, -55\,dBm), and high (HP, -15\,dBm) powers. The MP and HP traces are shifted in $\Delta Q_i$ by 5 and 10 for better distinction. (c) The relationship between interleaved measurements at different powers is illustrated using the signal coherence (i.e., normalized cross-spectral density) for each pair of powers. Focusing on low frequencies, the LP-MP pair shows greater correlation compared to LP-HP as the source of fluctuations at low and higher powers is different due to TLS saturation. The MP-HP pair similarly exhibits little coherence for the same reason.}
    \label{fig:interleaved}
\end{figure}

\begin{figure}[b!]
\includegraphics[width=\hsize]{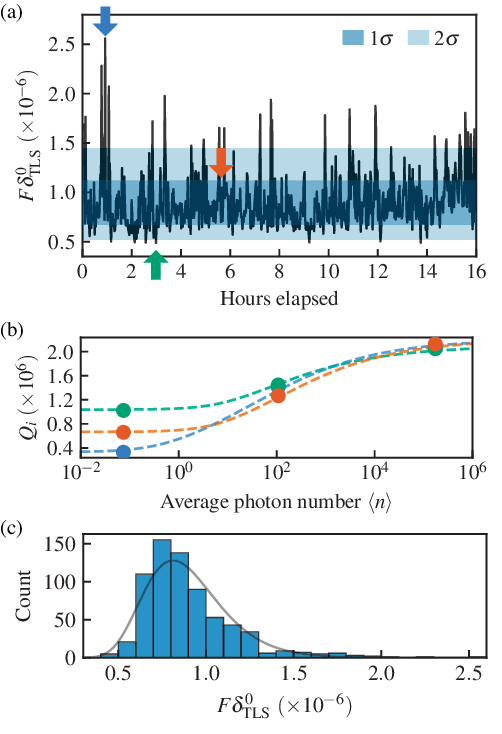}
    \caption{Fluctuations of the effective TLS loss tangent $F \delta^0_{\mathrm{TLS}}$. (a) Effective TLS loss tangent over 16 hours estimated from interleaved LP and HP measurements as $F \delta^0_{\mathrm{TLS}} \simeq 1/Q_{\text{LP}} - 1/Q_{\text{HP}}$. (b) Quality factors at LP, MP and HP, taken at instances of time shown by the arrows of the corresponding colors in (a). This illustrates how the TLS loss model curve varies over time, with the dashed lines being approximate fits to Eq.~\eqref{eq:TLS_loss} to serve as guides to the eye. (c) Distribution of $F \delta^0_{\mathrm{TLS}}$ for the data shown in (a), demonstrating the large spread in loss tangent. We obtain the $1\sigma$ and $2\sigma$ regions shown in (a) (shaded in blue) by fitting the histogram to a log-normal distribution (solid line).}
    \label{fig:FdTLS-fluctuations}
\end{figure}

We further study correlations in fluctuations at different powers by performing fast interleaved measurements over 16 hours at LP, MP, and HP. We used a JPA for the lower two powers to achieve average measurement times of 38\,s and 9\,s per point versus 10\,s for the high power measurements. The results notably show that the LP-MP pair exhibits strong correlations, in terms of both their temporal behavior and magnitude (see Fig.~\ref{fig:interleaved}). This is consistent with a saturable loss process, such as that associated with TLS. Namely, we expect correlations between $Q_i$ measurements at low and medium power, as TLS is the main source of fluctuations and the dominant loss channel at both power levels. This is further illustrated with the normalized cross-spectral density, or signal coherence (see Appendix E), showing strong correlation for the LP-MP pair especially at low frequencies. On the other hand, we observe practically no correlations for LP-HP at low frequencies since they are dominated by separate dissipation channels.

With the same interleaved measurements, we additionally track the effective TLS loss tangent $F \delta^0_{\mathrm{TLS}}$ from the quality factors at low ($Q_\text{LP}$) and high ($Q_\text{HP}$) powers using $F \delta^0_{\mathrm{TLS}} \simeq 1/Q_{\text{LP}} - 1/Q_{\text{HP}}$ (see Fig. \ref{fig:FdTLS-fluctuations}). This expression holds for low temperatures ($\hbar\omega \gg k_B T$ in Eq.~\eqref{eq:TLS_loss}) and for powers within the two plateaus of the S curve. We observe large variations of $F \delta^0_{\mathrm{TLS}}$ with $F \delta^0_{\mathrm{TLS}} = (9.0 \pm 2.2) \times 10^{-7}$ (Fig.~\hyperref[fig:FdTLS-fluctuations]{\ref*{fig:FdTLS-fluctuations}(a)}). In Fig.~\hyperref[fig:FdTLS-fluctuations]{\ref*{fig:FdTLS-fluctuations}(b)}, we further illustrate the temporal variations of the loss by plotting the approximate S curves at different time instances (as indicated by the arrows in Fig.~\hyperref[fig:FdTLS-fluctuations]{\ref*{fig:FdTLS-fluctuations}(a)}). We note that $Q_\text{LP}$ varies significantly between $3.3 \times 10^5$ and $1.0 \times 10^6$ while $Q_\text{HP}$ roughly stays fixed in time. This observation confirms that, even though the HP point was chosen as the highest power before the onset of nonlinearity in the resonator's response and may not lie within the high-power plateau, this choice would introduce only a small offset in the measurement of $F\delta^0_{\mathrm{TLS}}$. Fig.~\hyperref[fig:FdTLS-fluctuations]{\ref*{fig:FdTLS-fluctuations}(c)} shows that the distribution of $F \delta^0_{\mathrm{TLS}}$ is wide and has a pronounced tail toward lower loss values. We notice that the logarithm of the data appears normally-distributed. We perform a fit to a log-normal distribution and observe good agreement. By further investigating how the sample distribution depends on measurement time (see Appendix F), we find that low power measurements over a few hours should be sufficient to estimate the average TLS loss tangent with reasonable accuracy. However, averaging for such long times would inevitably result in the inability to characterize the width of the distribution, which may be of interest in materials study, for example. In that case, performing a series of measurements as described here would provide the distribution width.

The distribution of the TLS loss tangent may additionally inform our understanding of relaxation time $T_1$ in superconducting qubits dominated by TLS loss. Indeed, near-resonant TLS are known to cause recurring periods of strong relaxation, appearing as telegraphic dynamics in a $T_1$ time trace~\cite{Klimov_2018, Schlor_2019, Béjanin_2021}. Tuning the qubit frequency away from these features, the $T_1$ background is characterized by a near-Gaussian distribution of higher coherence. We hypothesize that, beyond decoherence caused by coupling to individual near-resonant TLS, the fluctuations in TLS loss tangent observed in this work may contribute to variations in the background $T_1$ in superconducting qubits. 

\begin{figure}[t]
\includegraphics[width=\hsize]{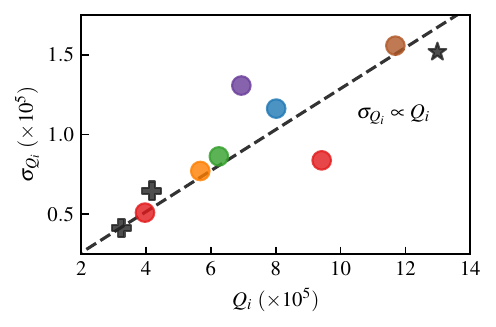}
    \caption{Standard deviation of the internal quality factor $\sigma_{Q_i}$ versus the average value at low power for various resonators. Our measurements results are shown as circles, where each color corresponds to a different device. The plus (star) signs are values obtained from Ref.~\citenum{Béjanin_2022} (Ref.~\citenum{Earnest_2018}). These results agree well with a linear model $\sigma_{Q_i} \propto Q_i$ (dashed line) obtained from considering an uniform TLS density of states with fluctuating couplings to the resonator~\cite{Neill_2013}. From it, we obtain an average standard deviation of $\sigma_{Q_i}/Q_i \simeq 13\%$.}
    \label{fig:sigmaQi_vs_Qi}
\end{figure}

To further elucidate our results, we compare the standard deviation of $Q_i$ with the mean value for low power measurements performed over several hours. We present results for various resonators measured over many cooldowns in addition to including results from other works \cite{Earnest_2018, Béjanin_2022} (see Fig. \ref{fig:sigmaQi_vs_Qi}). This illustrates that large fluctuations of the quality factor of superconducting resonators at low power appear to be universal across many devices. We observe good agreement with the linear relation $\sigma_{Q_i} \propto Q_i$. This is motivated by the use of an admittance model~\cite{Neill_2013} that assumes a uniform TLS density of states with fluctuating couplings to the resonator. From it, we obtain relative fluctuations of $\sigma_{Q_i}/Q_i \simeq 13\%$. We note that the proportionality constant is power-dependent: fitting a linear model to high power measurements (see Appendix H) similarly yields a reasonable agreement, with $\sigma_{Q_i}/Q_i \simeq 0.5\%$. The measurements in Fig. \ref{fig:sigmaQi_vs_Qi} were performed at similar but not identical low power levels, potentially explaining some deviations from the linear model.

In conclusion, we present results showing large internal quality factor fluctuations for distributed-element superconducting microwave resonators. These fluctuations are observed across resonators, chips and cooldowns. The variations in internal quality factor can be linked to fluctuations in the loss tangent induced by two-level systems in the device materials. We find that the mean and standard deviation of $Q_i$ obey a linear relationship. We demonstrate that performing time-averaged measurements over a few hours represents well the TLS loss tangent mean, while the standard deviation of the distribution is larger than the time-averaged fit uncertainty. Future work may help shed more light on the impact of different designs and materials and include the development of theoretical models to quantitatively explain our results. Further, we hypothesize that the TLS loss tangent fluctuations seen in this work are present in superconducting qubits as well. Future investigations may help identify a link between these fluctuations and the low-frequency background $T_1$ fluctuations observed in superconducting qubits dominated by TLS loss.

\section*{Acknowledgements}
We would like to thank NIST ERB reviewers Mark Keller, Adam Sirois, and Eva Gurra for valuable feedback. We acknowledge the Advanced Microwave Photonics Group (686.05) at NIST Boulder for providing the JPA used in this experiment. This material is based upon work supported by the U.S. Department of Energy, Office of Science, National Quantum Information Science Research Centers, Superconducting Quantum Materials and Systems Center (SQMS) under contract no. DE-AC02-07CH11359.

\section*{Author Declarations}
\subsection*{Conflict of Interest}
The authors have no conflicts to disclose.

\subsection*{Author Contributions}
\textbf{André Vallières:} conceptualization (lead); data curation (lead); formal analysis (lead); investigation (lead); methodology (lead); visualization (lead); writing - original draft preparation (lead); writing - review \& editing (lead). \textbf{Megan E. Russell:} conceptualization (supporting); formal analysis (supporting); methodology (supporting); writing - original draft preparation (supporting); writing - review \& editing (supporting). \textbf{Xinyuan You:} formal analysis (supporting); methodology (supporting); writing - review \& editing (supporting). \textbf{David A. Garcia-Wetten:} resources (equal); writing - review \& editing (supporting). \textbf{Dominic P. Goronzy:} resources (equal); writing - review \& editing (supporting). \textbf{Mitchell J. Walker:} resources (equal); writing - review \& editing (supporting). \textbf{Michael J. Bedzyk:} funding acquisition (supporting); supervision (supporting); writing - review \& editing (supporting). \textbf{Mark C. Hersam:} funding acquisition (lead); supervision (supporting); writing - review \& editing (supporting). \textbf{Alexander Romanenko:} supervision (supporting); writing - review \& editing (supporting). \textbf{Yao Lu:} methodology (supporting); writing - review \& editing (supporting). \textbf{Anna Grassellino:} funding acquisition (lead); supervision (supporting); writing - review \& editing (supporting). \textbf{Jens Koch:} funding acquisition (lead); methodology (supporting); supervision (lead); writing - review \& editing (supporting). \textbf{Corey Rae H. McRae:} conceptualization (supporting); formal analysis (supporting); methodology (supporting); supervision (lead); writing - original draft preparation (supporting); writing - review \& editing (supporting).

\subsection*{Data Availability Statement}
The data that support the findings of
this study are available from the
corresponding author upon reasonable
request.

\section*{References}

\begin{thebibliography}{33}
\bibitem [{Wel()}]{Welch_note}%
  \BibitemOpen
  \href@noop {} {\ }\bibinfo {note} {We used the \texttt{welch} function of the \texttt{scipy.signal} module in Python, applying a Hann window of length corresponding to 20\% of the total trace length, 16,384 FFT samples and no detrending. The cross-spectral density is obtained using the \texttt{csd} function.}\BibitemShut {Stop}%
\end{thebibliography}
\putbib
\end{bibunit}

\clearpage
\appendix

\renewcommand{\thefigure}{A.\arabic{figure}}
\setcounter{figure}{0}

\begin{bibunit}
\nocite{Kopas_2022}

\section{Chip design}
\label{sec:app-chip-design}
\begin{figure}[htb]
\includegraphics[width=0.8\hsize]{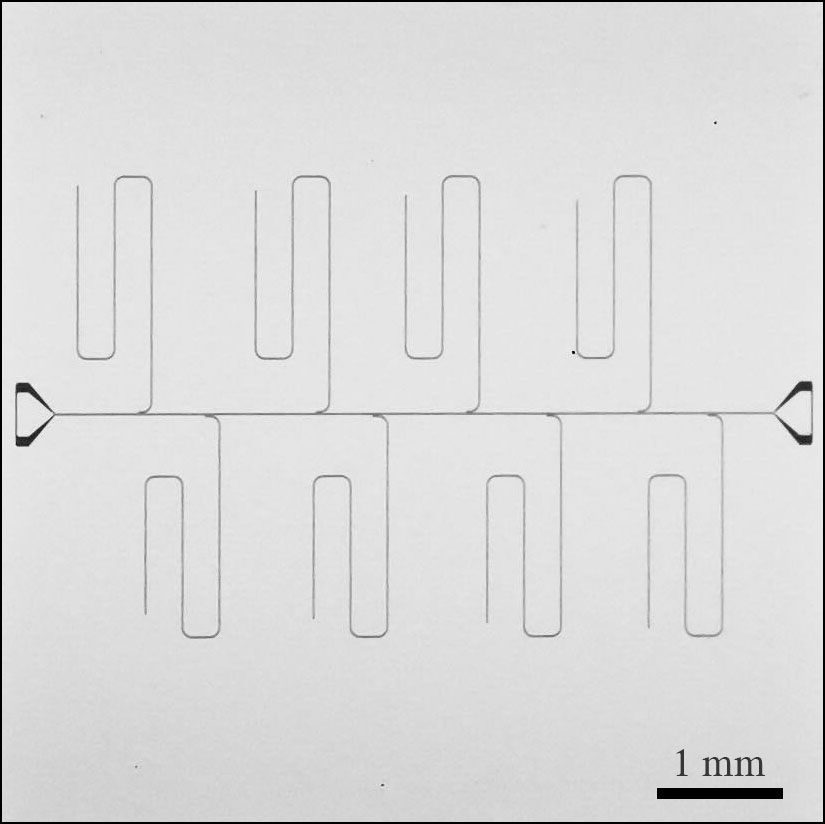}
    \caption{Micrograph of a sample chip consisting of eight inductively-coupled quarter-wavelength superconducting resonators.}
    \label{fig:chip-layout}
\end{figure}

Our coplanar waveguide resonator chips are based on the design described in Ref.~\citenum{Kopas_2022}. Room temperature, sputter-deposited Nb films on Si (001) and (111) were patterned into the coplanar waveguide resonators with optical lithography and SF$_6$-based reactive ion etching. Nb film thicknesses are (172 $\pm$ 5) nm. Each die is 7.5$\times$7.5 mm$^2$ and consists of eight, inductively coupled, quarter-wavelength resonators in the 4\,GHz to 8\,GHz range with target coupling quality factor of $5 \times 10^5$ to be close to critical coupling. An example chip design is shown in Fig.~\ref{fig:chip-layout}. The dies were rigidly fixed to Au-plated Cu packages with two small dots of GE varnish on each corner, and ground plane and transmission line bonds (not shown) were made using aluminum wedge bonding.

\begin{figure}[htb]
\includegraphics[width=0.9\hsize]{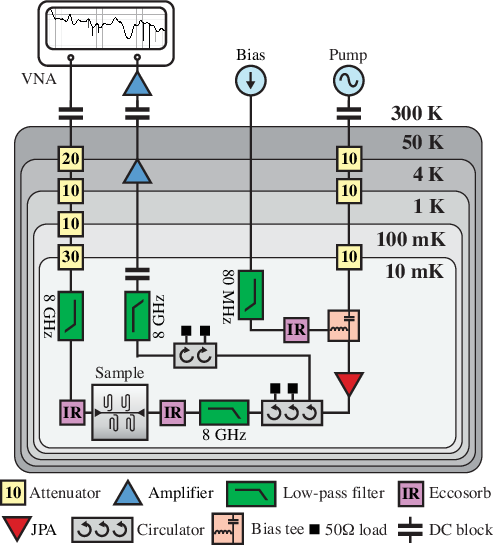}
    \caption{Wiring diagram for superconducting microwave resonator measurements. We use a VNA to probe our samples in the 4\,GHz to 8\,GHz range and a combination of microwave components to attenuate signals below single-photon level and amplify them above noise level at room temperature.}
    \label{fig:wiring-diagram}
\end{figure}

\section{Measurement setup}
\label{sec:app-measurement-setup}
Our measurement setup is shown in Fig. \ref{fig:wiring-diagram}. The input signal is first attenuated by a total of 70\,dB and filtered through an 8\,GHz low-pass filter then a commercial Eccosorb infrared (IR) filter. The output signal is amplified using a JPA anchored at the mixing chamber stage ($\sim$10\,mK), a high-electron mobility transistor (HEMT) amplifier at the 4\,K stage, and a room-temperature amplifier, with isolators and filters along the line to minimize noise returning to the sample and prevent saturation of the HEMT amplifier. We operate our JPA by tuning its frequency using a current source and pumping at roughly twice the signal frequency using an external microwave signal generator. Additionally, we are able to measure multiple samples in the cooldown by using cryogenic switches and sharing the input/output lines (not shown in schematic).

\begin{figure}[b!]
\includegraphics[width=\hsize]{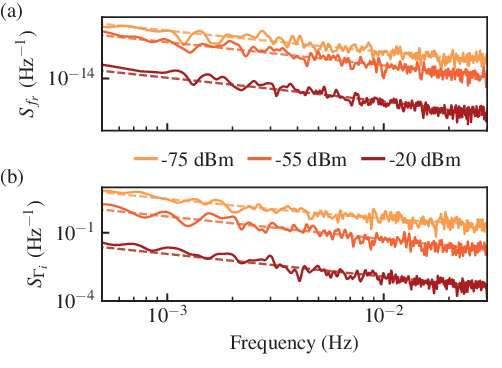}
    \caption{(a) Frequency and (b) decay rate noise spectra for various powers, where the legend refers to the power at the VNA. Fluctuations decrease with increasing power over the full range of frequencies, consistent with TLS saturation. The dashed lines are fits to a $1/f$ spectrum as guides to the eye.}
    \label{fig:fres_Gammai-spectra-vs-power} 
\end{figure}

We note that the room-temperature cabling and the stainless steel cables in the input line add approximately 20 dB of additional attenuation, bringing the total input attenuation from the VNA to roughly 90 dB. This value, together with the parameters obtained from Eq.~(1), is used to convert VNA power to average photon number $\langle n \rangle$~\cite{Baity_2024}. Since we are not using $\langle n \rangle$ as a comparative measure, the exact accuracy of the total input attenuation is not critical, and we acknowledge that it may be off by a few dB.

\section{Power dependence of $f_{\text{res}}$ and $\Gamma_i$ fluctuations}
As most of the prior works~\cite{Kumar_2008, Béjanin_2022} focused on resonance frequency ($f_r$) and internal decay rate ($\Gamma_i = 2\pi f_r/Q_i$) fluctuations, in Fig.~\ref{fig:fres_Gammai-spectra-vs-power} we show the spectral densities for these two variables at different powers. We estimate the spectral densities using Welch's method as explained in the main text. In agreement with these works, we see a decrease of the fluctuations for increasing power.

\begin{figure}[h]
\includegraphics[width=\hsize]{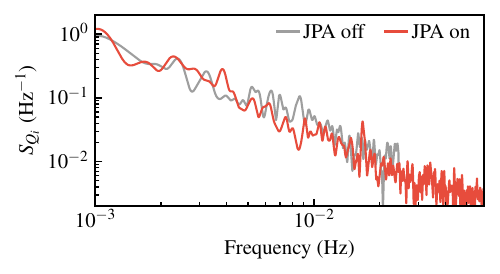}
    \caption{Internal quality factor noise spectra for measurements taken with and without the JPA. Each measurement was performed over 6 hours with measurement rates of 7\,s and 21\,s per sample, with and without the JPA respectively. The difference in spectra is negligible and could not explain the large fluctuations observed at low power.}
    \label{fig:with-without-JPA}
\end{figure}

\section{Added noise due to the JPA}
Since we use the JPA only at lower powers to achieve reasonable measurement rates, we must verify that the increased fluctuations at lower power are not caused by the amplification and that the JPA does not significantly increase $Q_i$ noise. To do so, in Fig.~\ref{fig:with-without-JPA}, we compare the internal quality factor noise spectra obtained from medium-power measurements with and without a JPA. We see that the only notable difference is that the noise floor is lower when the JPA is on as the amplification brings the signal above the white noise level at these frequencies. Otherwise, the spectra are basically the same and we can conclude that using the JPA does not negatively impact the measurements.

\section{Interleaved frequency measurement correlation}
\label{sec:app-interleaved-freq}
\begin{figure}[t]
\includegraphics[width=\hsize]{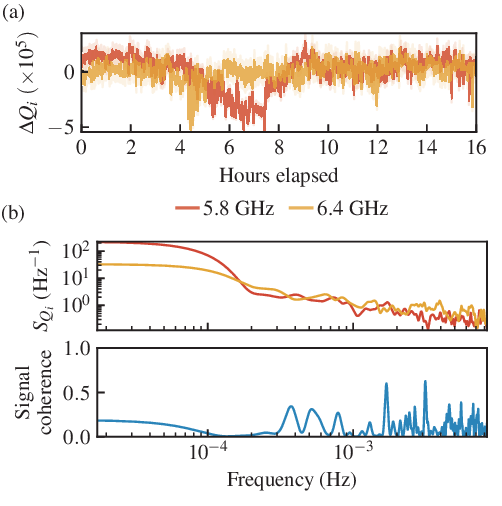}
    \caption{Interleaved resonator measurements of the internal quality factor over 16 hours with 30\,s per point. (a) Time traces of the deviation from the average $\Delta Q_i = Q_i - \langle Q_i \rangle$ for the two resonators at 5.8\,GHz and 6.4\,GHz. The measurements were performed at medium power (-55\,dBm at the VNA). (b) Internal quality factor noise spectra and signal coherence of the two resonators. The small signal coherence suggests that the fluctuations are local to each resonator.}
    \label{fig:coherence}
\end{figure}

To confirm that the fluctuations are not due to the measurement setup, but rather from local noise affecting the resonators, in Fig.~\ref{fig:coherence}, we show the signal coherence of two traces taken from interleaved measurements of the two resonators. For each resonator, $A$ and $B$, we first compute the normalized signal $(Q_i - \langle Q_i \rangle)/\langle Q_i \rangle$, where $\langle Q_i \rangle$ corresponds to the average value. Then, using Welch's method as explained in the main text~\cite{Welch_note}, we compute the signal coherence as $|S_{AB}|^2/(S_{AA}S_{BB})$, where $S_{AB}$ is the cross spectral density, and $S_{AA}$ or $S_{BB}$ the spectral density for resonators $A$ and $B$. We observe small correlations, indicating that the fluctuations are local to resonators rather than shared, as would be expected for amplifier noise, for example.

\begin{figure}[b!]
\includegraphics[width=\hsize]{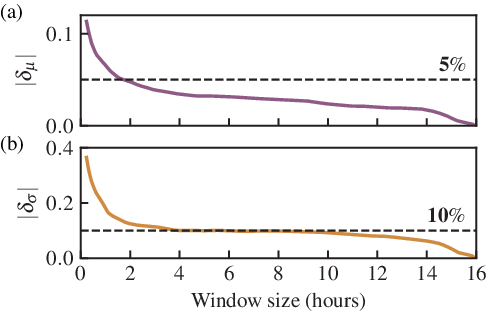}
    \caption{Relative deviation of the sample distribution parameters ($\mu$, $\sigma$) for the TLS loss tangent from their 16-hour values, as a function of window size. We observe that measuring over a few hours is necessary to achieve a target accuracy of 5\% for the mean and 10\% for the standard deviation.}
    \label{fig:FdTLS-dist-vs-time}
\end{figure}

\begin{figure*}[t!]
\includegraphics[width=\hsize]{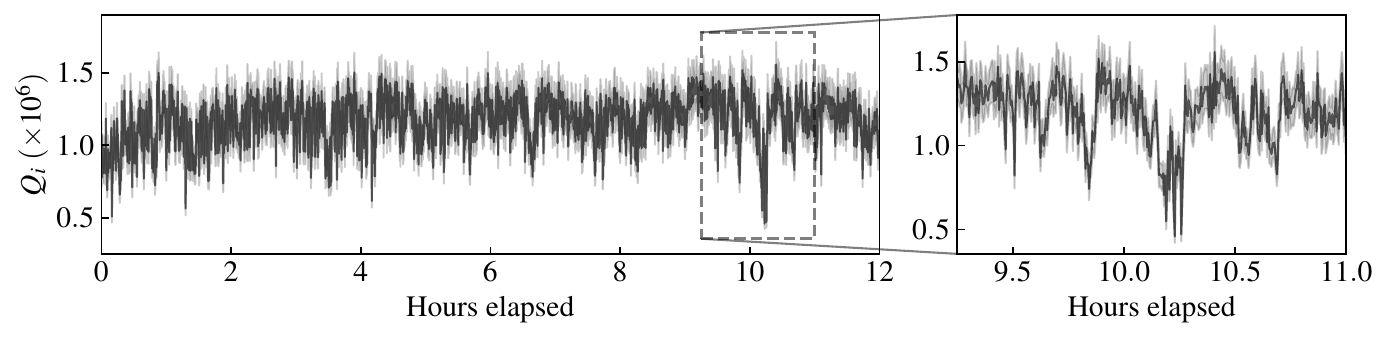}
    \caption{Time trace of the internal quality factor at low power. The zoomed-in region shows a significant drop of over 70\% from the maximum to minimum value in under 2 hours. The pale regions around the data represent fitting uncertainty.}
    \label{fig:full-width-trace}
\end{figure*}

\begin{figure}[b!]
\includegraphics[width=\hsize]{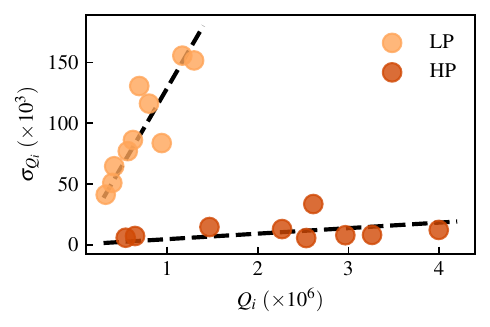}
    \caption{Standard deviation of the internal quality factor $\sigma_{Q_i}$ versus the average value of $Q_i$ at high (HP) and (LP) powers for various resonators (at base temperature $\sim$10\,mK). The dashed lines are fits to a linear model $\sigma_{Q_i} \propto Q_i$, from which we obtain average fluctuations of 0.5\% (13\%) for measurements at high (low) power.}
    \label{fig:sigmaQi_vs_Qi_HP}
\end{figure}

\section{TLS loss tangent distribution over time}
\label{sec:app-TLS-dist-vs-time}
We have seen that the effective TLS loss tangent $F\delta_{\mathrm{TLS}}^0$ follows a log-normal distribution, and can be modeled by an average $\mu$ and standard deviation $\sigma$. By investigating how these parameters change as we increase the measurement time, we can estimate for how long low-power measurements should be performed to obtain the TLS loss tangent with reasonable accuracy. To do so, we compute the sample distribution parameters ($\mu$ and $\sigma$) for overlapping measurement windows of various sizes. We aggregate the results by computing the average relative deviation of the mean and standard deviation from their respective long-term values, determined over a 16-hour period ($\mu_{16\text{h}}$ and $\sigma{16\text{h}}$). Specifically, the relative deviations are given by $\delta_\mu = (\mu - \mu_{16\text{h}})/\mu_{16\text{h}}$ and $\delta_\sigma = (\sigma - \sigma_{16\text{h}})/\sigma_{16\text{h}}$, and are shown in Fig.~\ref{fig:FdTLS-dist-vs-time}. For a target accuracy of 5\% in the mean (10\% in the standard deviation), we see that $\mu$ converges after about 2 hours, whereas for $\sigma$ it takes twice as long. From these results, we recommend measurements over a few hours at low power to obtain the mean and standard deviation of the TLS loss tangent to within reasonable accuracy.

\begin{figure}[h!]
\includegraphics[width=\hsize]{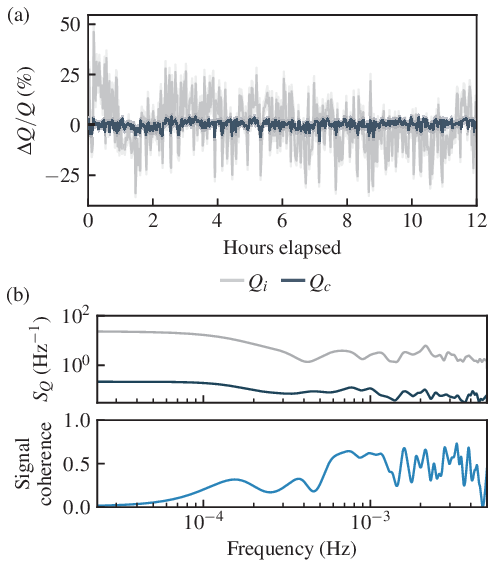}
    \caption{Fluctuations of the internal ($Q_i$) and coupling ($Q_c$) quality factors. (a) Relative change $Q_i$ and $Q_c$ with respect to their average over time, showing that $Q_c$ is stable in time. The pale regions around the data represent fitting uncertainty. (b) Noise spectra and signal coherence of $Q_i$ and $Q_c$.}
    \label{fig:Qc-independence}
\end{figure}

\section{Sample time trace}
In Fig.~\ref{fig:full-width-trace}, we show a 12-hour time trace of the internal quality factor at low power with a zoomed-in region of approximately 2 hours. We observe both slow and fast fluctuations, with occasional variations exceeding 70\% in under 2 hours, as illustrated on the right. These observations further demonstrate the need for extended measurements to accurately capture low-power $Q_i$ due to significant fluctuations.

\section{Linear trend in high power fluctuations}
\label{sec:app-high-power-sigmaQi-vs-Qi}
To further exemplify how the magnitude of fluctuations not only depends on $Q_i$ but additionally on the power, in Fig. \ref{fig:sigmaQi_vs_Qi_HP}, we have added high-power measurement data to Fig.~4 of the main text. The measurements were performed over many hours for various resonators across multiple cooldowns. We observe an acceptable agreement to a linear model, from which we obtain a relative standard deviation of 0.5\%. This suggests that as the number of photons in the resonator varies, the relation between $\sigma_{Q_i}$ and $Q_i$ stays roughly linear and can be useful in estimating the magnitude of fluctuations for a pair of power and $Q_i$.

\section{Investigation of $Q_c$ stability}
As explained in the main text, all reported measurements of $Q_i$ are obtained from fits to Eq.~(1), from which we also obtain the coupling quality factor, $Q_c$. As this value is fixed for a given resonator, confirming that it does not vary significantly over time helps validating our measurement procedure. From the low-power measurement results shown in Fig.~\hyperref[fig:Qc-independence]{\ref*{fig:Qc-independence}(a)}, we find that the average uncertainty and standard deviation of $Q_c$ are both around 2\%. There is also little correlation between $Q_i$ and $Q_c$ for long timescales (see Fig.~\hyperref[fig:Qc-independence]{\ref*{fig:Qc-independence}(b)}), further confirming that the observed fluctuations in $Q_i$ are not artificially induced by issues with fitting. The significant correlation for short timescales comes from the non-zero covariance between $Q_i$ and $Q_c$ that dominates for fluctuations on the order of the uncertainty.

\begin{figure}[t!]
\includegraphics[width=\hsize]{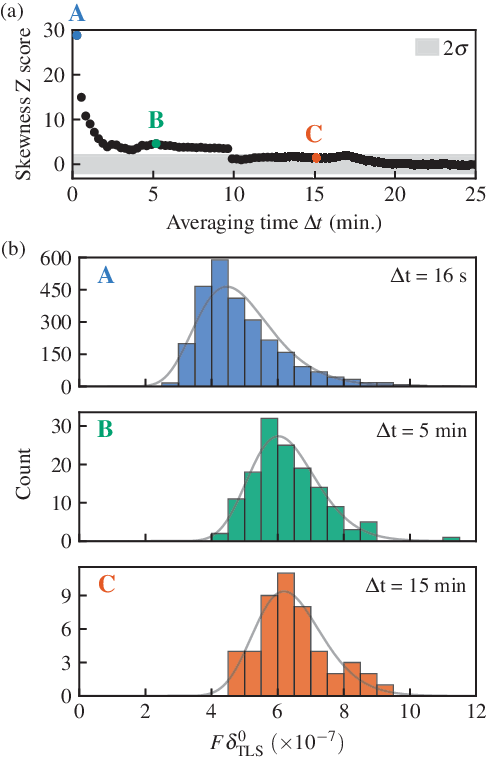}
    \caption{Normality testing of the $F\delta^0_{\mathrm{TLS}}$ distribution versus averaging time, for a total measurement time of 12 hours. Datasets of various averaging time are constructed by averaging the raw VNA traces and fitting the results to Eq. \eqref{eq:S21-hanger}. From this, we obtain the low power intrinsic quality factor $Q_{\text{LP}}$ and estimate the effective TLS loss tangent as $F \delta^0_{\mathrm{TLS}} \simeq 1/Q_{\text{LP}} - 1/Q_{\text{HP}}$ (high-power intrinsic quality factor $Q_{\text{HP}}$ obtained from prior measurements). (a) The skewness Z score is used as a metric of the normality of the distribution and is shown to progressively converge towards zero. The 95\% confidence interval is shown by the gray $2\sigma$ band ($Z=\pm 2$). (b) TLS loss tangent distributions at points A, B, and C, as shown in (a). The gray curves correspond to fits to a log-normal distribution.}
    \label{fig:FdTLS-dist-vs-meas-time}
\end{figure}

\section{Impact of measurement rate on the $F\delta^0_{\mathrm{TLS}}$ distribution}
$F\delta^0_{\mathrm{TLS}}$ is well described by a log-normal distribution when averaging time is short as in this study, but due to the central limit theorem, the distribution necessarily evolves into a more symmetric distribution as averaging time increases. This effect could potentially explain the discrepancy between the near-Gaussian background $T_1$ distribution in superconducting transmon qubits~\cite{Klimov_2018} and the log-normal distribution we observe in resonator TLS loss tangent.

We perform rapidly-sampled measurements of $S_{21}$ at low power, taken every 16 s over a 12-hour period, to construct datasets with varying averaging times. By averaging consecutive $S_{21}$ samples, we simulate averaging over longer times, up to an averaging time of $\Delta t \simeq 25$ minutes for 91 samples. From these datasets, we extract the low-power internal quality factor $Q_{\text{LP}}$ as a function of averaging time. By averaging the high power internal quality factor $Q_{\text{HP}}$ obtained from prior measurements, we estimate the effective TLS loss tangent as $F \delta^0_{\mathrm{TLS}} \simeq 1/Q_{\text{LP}} - 1/Q_{\text{HP}}$ and visualize its distribution as a function of averaging time (see Fig.~\ref{fig:FdTLS-dist-vs-meas-time}). The metric we use for normality testing is the skewness Z score computed using \texttt{scipy.stats}'s \texttt{skewtest} function, which measures the symmetry of the distribution. 

We note that at short averaging times the distribution is skewed, and progressively evolves into a more symmetric shape as averaging time increases. This is expected from the central limit theorem, where we can predict the skewness to decrease as $1/\sqrt{\Delta t}$. However, this does not describe the full behavior observed such as the apparent jump at $\Delta t = 10$ minutes. Nevertheless, we notice that the initial decrease in Z score plateaus at around $\Delta t = 2$ minutes, indicating that fast enough measurements of the loss can capture short moments of lower $Q_i$. This suggests that the $T_1$ background in superconducting qubits may look more skewed if sampled at a faster rate assuming the underlying physical mechanisms for fluctuations are similar.\\

\newpage

\section*{References}
\vspace{-1.5em}
\putbib
\end{bibunit}

\newpage

\end{document}